\documentclass[aip,pop,11pt,a4paper,preprint,amsmath,amssymb,groupedaddress]{revtex4-2}

\usepackage{bm}
\usepackage{graphicx}

\usepackage{times}
\usepackage{mathptmx}

\newcommand{\p}{\partial}

\newcommand{\ave}[1]{{\left<#1\right>}}

\DeclareMathOperator\erfc{erfc}

\newcommand{\rmd}{\text{d}}

\newcommand{\tauw}{\ensuremath{\langle{w_0}\rangle}}
\newcommand{\tauwxi}{\ensuremath{\langle{w_\xi}\rangle}}
\newcommand{\taustar}{\tau_*}
\newcommand{\gammastar}{\gamma_*}

\newcommand{\axi}{{a}}
\newcommand{\Axi}{{A}}
\newcommand{\Vxi}{{V}}

\newcommand{\tauxi}{\tau}
\newcommand{\Aave}{\langle{A}\rangle}

\newcommand{\aave}{\langle{a}\rangle}
\newcommand{\aoave}{\langle{a_0}\rangle}

\newcommand{\Vave}{\langle{V}\rangle}

\newcommand{\voave}{\langle{v_0}\rangle}

\newcommand{\wave}{\langle{w}\rangle}
\newcommand{\woave}{\langle{w_0}\rangle}

\newcommand{\Koave}{\langle{K_0}\rangle}
\newcommand{\Kxiave}{\langle{K_\xi}\rangle}
\newcommand{\Phiave}{\ensuremath{\ave{\Phi}}}
\newcommand{\Phirms}{\ensuremath{\Phi}_\text{rms}}

\newcommand{\taup}{\ensuremath{\tau_\shortparallel}}

\newcommand{\aomin}{\ensuremath{a_{0\,\text{min}}}}
\newcommand{\umax}{\ensuremath{u_\text{max}}}
\newcommand{\Xmax}{\ensuremath{X_\text{max}}}

\newcommand{\rmedian}{\ensuremath{r_\text{med}}}

\newcommand{\Eqref}[1]{Eq.~\eqref{#1}}
\newcommand{\Eqsref}[1]{Eqs.~\eqref{#1}}
\newcommand{\Figref}[1]{Fig.~\ref{#1}}
\newcommand{\Figsref}[1]{Figs.~\ref{#1}}
\newcommand{\Secref}[1]{Sec.~\ref{#1}}
\newcommand{\Secsref}[1]{Secs.~\ref{#1}}
\newcommand{\Tabref}[1]{Tab.~\ref{#1}}

\begin{document}

\title{Stochastic modeling of blob-like plasma filaments in the scrape-off layer: Time-dependent velocities and pulse stagnation}

\author{O.~Paikina}
\email{olga.paikina@uit.no}

\author{J.~M.~Losada}
\email{juan.m.losada@uit.no}

\author{A.~Theodorsen}
\email{audun.theodorsen@uit.no}

\author{O.~E.~Garcia}
\email{odd.erik.garcia@uit.no}

\affiliation{Department of Physics and Technology, UiT The Arctic University of Norway, N-9037 Troms{\o}, Norway}

\date{\today}

\begin{abstract}
A stochastic model for a super-position of uncorrelated pulses with a random distribution of and correlations between amplitudes and velocities is analyzed. The pulses are assumed to move radially with fixed shape and amplitudes decreasing exponentially in time due to linear damping. The pulse velocities are taken to be time-dependent with a power law dependence on the instantaneous amplitudes, as suggested by blob velocity scaling theories. In accordance with experimental measurements, the pulse function is assumed to be exponential and the amplitudes are taken to be exponentially distributed. As a consequence of linear damping and time-dependent velocities, it is demonstrated that the pulses stagnate during their radial motion. This makes the average pulse waiting time increase radially outwards in the scrape-off layer of magnetically confined plasmas. In the case that pulse velocities are proportional to their amplitudes, the mean value of the process decreases exponentially with radial coordinate, similar to the case when all pulses have the same, time-independent velocity. The profile e-folding length is then given by the product of the average pulse velocity and the parallel transit time. Moreover, both the average pulse amplitude and the average velocity are the same at all radial positions due to stagnation of slow and small-amplitude pulses. In general, an increasing average pulse velocity results in a flattened radial profile of the mean value of the process as well as a higher relative fluctuation level, strongly enhancing plasma--surface interactions.
\end{abstract}

\maketitle

\clearpage

\section{Introduction}\label{sec.intro}

The boundary region of magnetically confined plasmas is commonly in an inherently fluctuating state with an order unity relative fluctuation level of the plasma parameters.\cite{wootton_fluctuations_1990,nedospasov_edge_1992,endler_turbulent_1999,dippolito_blob_2004,carreras_plasma_2005,zweben_edge_2007,naulin_turbulent_2007,krasheninnikov_recent_2008,garcia_blob_2009,dippolito_convective_2011} These fluctuations are generally attributed to magnetic field-aligned plasma filament structures that move radially outwards to the main chamber walls.\cite{dippolito_blob_2004,carreras_plasma_2005,zweben_edge_2007,naulin_turbulent_2007,krasheninnikov_recent_2008,garcia_blob_2009,dippolito_convective_2011} These structures are localized in the drift plane perpendicular to the magnetic field. Moreover, they have excess pressure compared to the time-average plasma state, and are therefore commonly referred to as blobs.\cite{zweben_search_1985,zweben_visible_1989,krasheninnikov_scrape_2001,bian_blobs_2003,grulke_radially_2006} The radial motion of blob-like filaments leads to flattened profiles of particle density and temperature in the scrape-off layer (SOL) and causes detrimental interactions with limiter structures, wave antennas and the main chamber walls.\cite{asakura_sol_1997,labombard_cross-field_2000,labombard_particle_2001,boedo_transport_2001,lipschultz_investigation_2002,antar_universality_2003,boedo_transport_2003,rudakov_far_2005,whyte_magnitude_2005,garcia_interchange_2006,lipschultz_plasma-surface_2007,garcia_fluctuations_2007,terry_scrape-off_2007,carralero_experimental_2014,boedo_edge_2014,vinello_modification_2017,carralero_study_2017,kube_statistical_2019,vianello_scrape-off_2020,stagni_dependence_2022}

Theoretical analysis and numerical simulations provide detailed insight to the blob dynamics, their velocity scaling and the correlation of their parameters.\cite{garcia_mechanism_2005,madsen_influence_2011,kube_effect_2012,wiesenberger_radial_2014,olsen_temperature_2016,pecseli_solvable_2016,held_influence_2016,wiesenberger_unified_2017,held_beyond_2023,garcia_radial_2006,kube_velocity_2011,manz_filament_2013,omotani_effects_2015,walkden_dynamics_2016,dippolito_cross-field_2002,myra_collisionality_2006,theiler_cross-field_2009,kube_velocity_2011,easy_three_2014,halpern_three-dimensional_2014,omotani_effects_2015,easy_investigation_2016,walkden_dynamics_2016} In all cases, the non-linear advection leads to a steep front and a trailing wake that is well described by a two-sided exponential pulse function.\cite{bian_blobs_2003,garcia_mechanism_2005} It is generally found that the radial blob velocity increases with the amplitude. Two distinct regimes have been identified. In the so-called inertial regime where sheath resistance is negligible, the velocity scales as the square root of the amplitude.\cite{garcia_mechanism_2005,madsen_influence_2011,kube_effect_2012,wiesenberger_radial_2014,olsen_temperature_2016,pecseli_solvable_2016,held_influence_2016,wiesenberger_unified_2017,held_beyond_2023,garcia_radial_2006,kube_velocity_2011,manz_filament_2013,omotani_effects_2015,walkden_dynamics_2016} In the sheath dissipative regime where inertia is less dominant, the velocity is proportional to the amplitude.\cite{dippolito_cross-field_2002,garcia_radial_2006,myra_collisionality_2006,theiler_cross-field_2009,kube_velocity_2011,manz_filament_2013,easy_three_2014,halpern_three-dimensional_2014,omotani_effects_2015,easy_investigation_2016,walkden_dynamics_2016}

In order to describe the cross-field transport by blob-like structures, a statistical description is necessary, taking into account the distribution of and correlations between the blob amplitudes and velocities. A statistical framework has been developed that describes both the time-averaged radial profiles and the fluctuations resulting from a super-position of uncorrelated pulses moving radially outwards.\cite{garcia_stochastic_2012,kube_convergence_2015,theodorsen_level_2016,theodorsen_statistical_2017,garcia_auto-correlation_2017,theodorsen_level_2018,theodorsen_probability_2018,garcia_stochastic_2016,losada_stochastic_2023,losada_stochastic_2024,militello_scrape_2016,militello_relation_2016,militello_two-dimensional_2018,ahmed_reconstruction_2023,losada_three-point_2024}
Both the underlying assumptions of this model as well as its predictions have been found to compare favorably with experimental measurements on various tokamak devices. In particular, it has been clearly demonstrated that single-point measurements record the blob structures as bursts with a two-sided exponential shape and an exponential distribution of peak amplitudes and the waiting times between them.\cite{garcia_burst_2013,garcia_intermittent_2013,kube_fluctuation_2016,theodorsen_scrape-off_2016,garcia_sol_2017,theodorsen_relationship_2017,walkden_statistical_2017,walkden_interpretation_2017,kube_intermittent_2018,kuang_plasma_2019,bencze_characterization_2019,kube_comparison_2020,zurita_stochastic_2022,zweben_temporal_2022,ahmed_strongly_2023}

Based on these statistical properties, the stochastic modeling shows that for the simplest case where all pulses have the same velocity, the fluctuations are Gamma distributed with the shape parameter given by the ratio of the average pulse duration and waiting times.\cite{garcia_stochastic_2012,kube_convergence_2015,theodorsen_level_2016,theodorsen_statistical_2017,garcia_auto-correlation_2017,theodorsen_level_2018,theodorsen_probability_2018,garcia_stochastic_2016,losada_stochastic_2023,losada_stochastic_2024} The resulting mean value is given by the product of the average amplitude and the shape parameter. Moreover, the mean amplitude has an exponential decreases with radial position with the e-folding length given by the product of the radial pulse velocity and the linear damping time due to losses along magnetic field lines in the SOL.\cite{garcia_stochastic_2016,losada_stochastic_2023,losada_stochastic_2024,militello_scrape_2016,militello_relation_2016} In the case of a broad distribution of pulse velocities, the downstream process is dominated by the fast pulses, which have short radial transit times and less amplitude reduction due to the linear damping.\cite{losada_stochastic_2023} When the pulse amplitudes and velocities are correlated, the radial variation of the mean value is close to the case of a degenerate distribution of pulse velocities but with larger relative fluctuation amplitudes and more intermittent fluctuations.\cite{losada_stochastic_2024}

Due to the linear damping, the amplitudes decrease exponentially in time as the pulses move radially outwards. This contribution considers the case where the pulse velocity has a power law dependence on the instantaneous pulse amplitude, leading to time-dependent velocities. This is demonstrated to result in stagnation of pulses and thereby a radial decrease of the rate of pulses. Closed form expressions are derived for the probability distribution and radial variation of pulse amplitudes, velocities and durations in the case of an exponential pulse function. Radial variation of the mean value, relative fluctuation level and the skewness and flatness moments are obtained in the cases of a square root and a linear velocity dependence on the instantaneous amplitude. In the latter case, appropriate for the sheath dissipative regime, the mean value of the process is exactly the same as when all pulses have the same, time-independent velocity. In this case, both the average pulse amplitude and velocity are the same at all radial positions. However, the fast and large-amplitude pulses dominate the process, resulting in large relative fluctuation levels and enhanced intermittency. In magnetically confined plasmas, this will result in strong fluctuation enhancement of plasma--surface interactions.

This paper is organized as follows. In \Secref{sec.model} we present the statistical framework, describing the plasma fluctuations due to a super-position of uncorrelated pulses with time-dependent velocities. The special case of time-independent velocities are reviewed and some new results presented for correlated pulse amplitudes and velocities. In \Secref{sec.t-dv} pulse velocities which have a power law dependence on the instantaneous pulse amplitudes are considered. This is shown to result in stagnation of pulses, multiple temporal scales for the pulse duration, and radial variation of the rate of pulses. In \Secref{sec.profiles} the radial variation of the lowest order moments for the process is discussed, considering both the square root and the linear scaling of velocity on amplitude. A discussion of the results and the main conclusions are presented in \Secsref{sec.disc} and \ref{sec.conclusions}, respectively. For completeness, the stationarity of the process is addressed in an appendix.

\section{Stochastic model}\label{sec.model}

In the first part of this section, we review the stochastic model describing a super-position of uncorrelated pulses, establishing notation and defining the reference case where all pulses have the same, time-independent velocity. In the last part, new results will be presented for the case where pulse velocities have a power law dependence on the initial pulse amplitudes.

\subsection{Super-position of pulses}

Consider a super-position of $K_0$ uncorrelated pulses moving radially along the $x$-axis described by
\begin{equation} \label{eq.PhiK}
    \Phi_{K_0}(x,t) = \sum_{k=1}^{K_0} \phi_k(x,t-s_{0k}) ,
\end{equation}
where the pulse $\phi(x,t-s_0)$ is located at the reference position $x=0$ at the arrival time $s_0$. Here and in the following,  for the simplicity of notation, the $k$-subscript on random variables is suppressed except when summation is explicit. A zero subscript on any variable indicates that it is specified at the reference position. The evolution of each pulse is determined by the advection-dissipation equation
\begin{equation}\label{eq.dphikdt}
    \frac{\p\phi}{\p t} + V\,\frac{\p \phi}{\p x} + \frac{\phi}{{\taup}} = 0 .
\end{equation}
Here $V$ is the pulse velocity and $\taup$ is a constant linear damping time assumed to be the same for all pulses. The pulse velocity $V(t)$ will in general be taken to be a function of time and may depend on other pulse parameters.

All pulses are assumed to have the same radial variation so that the pulse function is given by the initial condition
\begin{equation}
    \phi(x,0) = a_{0} \varphi\left( \frac{x}{\ell} \right) ,
\end{equation}
where $\ell$ is the pulse size and $a_{0}$ is the pulse amplitude at the reference position. The general solution of the advection equation~\eqref{eq.dphikdt} can thus be written as
\begin{equation}\label{eq.phik}
    \phi(x,t) = A(t) \varphi\left( \frac{x-X(t)}{\ell} \right) ,
\end{equation}
where the pulse position $X(t)$ satisfies $\text{d}X/\text{d}t=V$ and the pulse amplitude evolution is determined by
\begin{equation}\label{eq.dAdt}
    \frac{\text{d}A}{\text{d}t} + \frac{A}{\taup} = 0 .
\end{equation}
The solution for the pulse amplitude is given by
\begin{equation}\label{eq.amplitude}
    A(t) = a_{0} \exp \left( - \frac{t}{\taup} \right) ,
\end{equation}
revealing exponential decrease in time due to the linear damping in \Eqsref{eq.dphikdt} and \eqref{eq.dAdt}. The radial position of the pulse at any given time $t$ is given by straight forward integration,
\begin{equation}\label{eq.Xk}
    X(t) = \int_{0}^{t} \rmd t'\, V(t') ,
\end{equation}
for any pulse velocity function $V(t)$, which in general may depend on $A$ and $\ell$.

With these specifications, the process given by \Eqref{eq.PhiK} can be written as
\begin{equation}\label{eq.PhiK.X}
   \Phi_{K_0}(x,t) = \sum_{k=1}^{K_0} a_{0k}\exp\left( -\frac{t-s_{0k}}{\taup} \right)\varphi\left( \frac{x-X_k(t-s_{0k})}{\ell_k} \right) .
\end{equation}
In the following, we will assume the size $\ell$ to be the same for all pulses and consider the case of a one-sided exponential pulse function,
\begin{equation}\label{eq.phiexp}
    \varphi(\theta) = \Theta(-\theta)\exp{(\theta)} ,
\end{equation}
where $\Theta$ denotes the unit step function,
\begin{equation}\label{eq.unitstep}
    \Theta(\theta) = \begin{cases}
     1, & \theta \geq 0, \\
     0, & \theta < 0 .
\end{cases}
\end{equation}
As a reference case, we will furthermore consider exponentially distributed pulse amplitudes at the reference position $x=0$, which for $a_{0}>0$ is given by
\begin{equation}\label{eq.pdfaoexp}
    \langle{a_0}\rangle P_{a_0}(a_0) = \exp{\left( -\frac{a_0}{\langle{a_0}\rangle} \right)} .
\end{equation}
Here and in the following, angular brackets denote the ensemble average of a random variable over all its arguments.

The pulse arrival times $s_0$ at the reference position are in the following assumed to be uniformly distributed on an interval of duration $T$, that is, their probability distribution function is $P_{s_0}(s_0)=1/T$ for $s_0\in[-T/2,T/2]$. The pulse arrival times are assumed to be independent of all other pulse parameters. With these assumptions, the probability that there are exactly $K_0$ pulse arrivals at $x=0$ during any interval of duration $T$ is given by the Poisson distribution
\begin{equation} \label{eq.poisson}
    P_{K_0}(K_0;T) = \frac{1}{K_0!}\left(\frac{T}{\tauw}\right)^{K_0}\exp\left(-\frac{T}{\tauw} \right) ,
\end{equation}
where $\tauw$ is the average pulse waiting time at the reference position. The average number of pulses in realizations of duration $T$ is $\Koave=T/\tauw$. From the Poisson distribution it follows that the waiting time between two subsequent pulses is exponentially distributed. As will be seen in \Secref{sec.t-dv}, time-dependent velocities implies a radial variation in the average pulse waiting times. However, before discussing the effect of time-dependent velocities, we will review the case where the pulses are uncorrelated and move with time-independent velocities.\cite{garcia_stochastic_2016,losada_stochastic_2023,losada_stochastic_2024,militello_scrape_2016,militello_relation_2016}

\subsection{Time-independent velocities}\label{sec.tiv}

In the special case that the pulse velocities are time-independent, $V(t)=v_0$, but may be randomly distributed, the radial pulse position from \Eqref{eq.Xk} is given by $X(t)=v_0t$, and the solution for the pulse function can be written as
\begin{equation}
    \phi(x,t-s_0) = a_0\exp\left(-\frac{t-s_0}{\taup}\right)\varphi\left(\frac{x-v_{0}(t-s_0)}{\ell}\right)
\end{equation}
For any given radial position $\xi$, the exponential decrease due to linear damping combines with the one-sided exponential pulse function, and the process recorded at this location is given by
\begin{equation}\label{eq.Phixi}
    \Phi_{K_0}(\xi,t) = \sum_{k=1}^{K_0} a_{\xi k}\varphi\left( \frac{t-s_{\xi k}}{\tau_{0k}} \right) ,
\end{equation}
where the pulse arrival time at the position $\xi$ is $s_\xi=s_0+\xi/v_0$, the pulse amplitude decreases radially outwards due to the linear damping,
\begin{equation}\label{eq.axiexp}
    a_{\xi} = a_0\exp\left( -\frac{\xi}{v_0\taup} \right) ,
\end{equation}
and the pulse duration is the harmonic mean of the radial transit time and the linear damping time,
\begin{equation}\label{eq.tau}
    \tau_0 = \frac{\taup\ell}{v_0\taup+\ell} .
\end{equation}
In this case, the duration of any pulse will be the same at all radial positions. When the pulse velocities have a random distribution, the distribution of pulse amplitudes $a_\xi$ at $\xi\neq0$ will be different from the ones specified at the reference position. The amplitude of slow pulses with long radial transit times will decrease substantially with radial position, and the downstream process will be dominated by the fast pulses. The amplitude distribution will be discussed further in \Secref{sec:corrav}.

The average pulse duration is obtained by integrating over the probability distribution for pulse sizes and velocities. A variation of the pulse velocities also implies a distribution of pulse durations as described by \Eqref{eq.tau}. When all pulses have the same size, the pulse duration distribution is related to the velocity distribution by the standard rules for the transformation of random variables,
\begin{equation}\label{eq.ptau}
    P_{\tau_0}(\tau_0) = \frac{\ell}{\tau_0^2}\,P_{v_0}\left( \frac{\ell}{\tau_0}-\frac{\ell}{\taup} \right) .
\end{equation}
Since the pulse duration is inversely proportional to the velocity, the average duration is dominated by the slow pulses in the case of a random velocity distribution. 

When there is a distribution of pulse velocities, the amplitudes and durations will be correlated downstream even if they are independent at the reference position. Fast pulses have short radial transit times and less amplitude reduction due to linear damping. As a result, the relative fluctuation level and the skewness and flatness moments increase radially outwards, as will be shown in \Secref{sec:corrav}.

The lowest-order statistical moments of the process can be derived from the cumulants $\kappa_n$, which
are the coefficients in the expansion of the logarithm of the characteristic function. The mean value is given by the first order cumulant, $\kappa_1=\Phiave$. The variance is given by the second order cumulant, $\kappa_2=\Phi_{\text{rms}}^2=\ave{(\Phi-\ave{\Phi})^2}$. The third and fourth order centered moments $\mu_n=\ave{(\Phi-\ave{\Phi})^n}$ are related to the cumulants by the relations $\mu_3=\kappa_3$ and $\mu_4=\kappa_4+3\kappa_2^2$. Thus, the skewness and flatness moments are defined respectively by
\begin{subequations}
\begin{align}
    S_{\Phi} & =\frac{\langle{(\Phi-\Phiave)^3}\rangle}{\Phi_\text{rms}^3} = \frac{\kappa_3}{\kappa_2^{3/2}} ,
    \\
    F_{\Phi} & = \frac{\langle{(\Phi-\Phiave)^4}\rangle}{\Phi_\text{rms}^4} - 3 = \frac{\kappa_4}{\kappa_2^{2}} .
\end{align}
\end{subequations}
When the process $\Phi$ has a normal distribution, both of the two latter moments vanish. For an exponential pulse function and time-independent pulse velocities, the cumulants are given by\cite{losada_stochastic_2023}
\begin{equation}
    \kappa_n(x) = \frac{1}{n\tauw} \ave{ a_0^n\tau_0\exp\left( - \frac{nx}{v_0\taup} \right) } ,
\end{equation}
where the pulse duration is given by \Eqref{eq.tau}. Here the averaging is to be taken over the pulse amplitudes, sizes and velocities, which in general may all be randomly distributed and correlated.

\subsection{Filtered Poisson process}\label{sec.fpp}

In the case that all pulses have the same velocity and size,
the cumulants are given by\cite{losada_stochastic_2023}
\begin{equation}\label{eq.kappan.degenerate}
    \kappa_n(x) = \gammastar (n-1)! \ave{a_0}^n \exp{\left( - \frac{nx}{\voave\taup} \right) } ,
\end{equation}
where we have defined the intermittency parameter $\gammastar$ as the ratio of the average pulse duration and waiting times,
\begin{equation}
    \gammastar = \frac{\taustar}{\tauw} ,
\end{equation}
where the average pulse duration is given by
\begin{equation}\label{eq.taustar}
    \taustar = \frac{\taup\ell}{\voave\taup+\ell} .
\end{equation}
The intermittency parameter determines the degree of pulse overlap. The cumulants given by \Eqref{eq.kappan.degenerate} describe a Gamma distribution with shape parameter $\gammastar$ and scale parameter given by the radial pulse amplitude profile corresponding to \Eqref{eq.axiexp},
\begin{equation}\label{eq.aavex}
    \aave = \langle{a_0}\rangle\exp{\left( - \frac{x}{\voave\taup} \right)} .
\end{equation}
The Gamma probability density function for positive $\Phi$ is thus given by\cite{garcia_stochastic_2012,garcia_stochastic_2016}
\begin{equation}
    \langle{a}\rangle P_{\Phi}(\Phi;x) = \frac{1}{\Gamma(\gammastar)} \left(\frac{\Phi}{\langle{a}\rangle}\right)^{\gammastar-1}\exp{\left(-\frac{\Phi}{\langle{a}\rangle}\right)} ,
\end{equation}
where $\Gamma$ denotes the Gamma function defined by
\begin{equation}
    \Gamma(\gamma) = \int_0^{\infty} \text{d}\zeta\,\zeta^{\gamma-1}\exp{(-\zeta)} .
\end{equation}
The mean value of the process is just the first order cumulant,\cite{garcia_stochastic_2016}
\begin{equation}
    \ave{\Phi}(x) = \gammastar \langle{a_0}\rangle \exp{\left( - \frac{x}{\voave\taup} \right)} .
\end{equation}
The lowest order normalized moments are in this case radially constant and determined solely by the intermittency parameter,\cite{losada_stochastic_2023}
\begin{equation}
    \frac{\Phi_\text{rms}}{\ave{\Phi}} = \frac{1}{\gammastar^{1/2}} ,
    \quad
    S_\Phi = \frac{2}{\gammastar^{1/2}} ,
    \quad
    F_\Phi = \frac{6}{\gammastar} .
\end{equation}
The mean value $\Phiave$ decreases exponentially with radius with an e-folding length $\voave\taup$, given by the product of the radial pulse velocity and the linear damping time. The prefactor $\gammastar\langle{a_0}\rangle$ is given by the product of the average pulse amplitude at the reference position and the ratio of the average pulse duration and waiting times. At any given position, this process describe a super-position of uncorrelated pulses with fixed duration and is commonly referred to as a filtered Poisson process. This defines the reference case in order to compare with a distribution of pulse velocities.

\subsection{Correlated amplitudes and velocities}\label{sec:corrav}

As described in \Secref{sec.intro} and in more detail in Ref.~\onlinecite{losada_stochastic_2024}, a particularly interesting case to consider is pulse velocities correlated with amplitudes through a power law dependence. At the reference position this is given by
\begin{equation}\label{eq.v0a0powerlaw}
    \frac{v_0}{\voave} = c_v \left( \frac{a_0}{\aoave} \right)^\alpha ,
\end{equation}
with the proportionality constant $c_v=\aoave^\alpha/\langle{a_0^\alpha}\rangle$ determined by the amplitude distribution. For the exponential amplitude distribution defined by \Eqref{eq.pdfaoexp}, the proportionality constant is given by $c_v(\alpha)=1/\Gamma(1+\alpha)$. It follows in particular that $c_v(0)=1$, $c_v(1/2)=2/\pi^{1/2}$ and $c_v(1)=1$. In the limit $\alpha\rightarrow0$, the pulse velocity becomes independent of the amplitude and has a degenerate distribution, that is, all pulses move with the same, time-independent velocity. The stochastic model then reduces to the filtered Poisson process discussed in the previous subsection. When $\alpha=1$, there is a linear relationship between $v_0$ and $a_0$, and the velocity distribution is the same as the amplitude distribution at the reference position.

Given the probability density function of the pulse amplitudes, $P_{a_0}$, the velocity distribution $P_{v_0}$ follows from the standard rules for transformation of random variables. For the power law relationship given by \Eqref{eq.v0a0powerlaw}, the velocity distribution is
\begin{equation}
    \voave P_{v_0} (v_0;\alpha) = \frac{\aoave}{\alpha c_v} \left(\frac{v_0}{c_v\voave}\right)^{\frac{1-\alpha}{\alpha}} P_{a_0}\left( \aoave \left( \frac{v_0}{c_v\voave} \right)^{\frac{1}{\alpha}} \right) .
\end{equation}
In the reference case of an exponential distribution of pulse amplitudes at the reference position, the probability density function of the pulse velocities is for $v_0>0$ given by
\begin{equation}
    \voave P_{v_0}(v_0;\alpha) = \frac{1}{\alpha c_v} \left( \frac{v_0}{c_v\voave}\right)^{\frac{1-\alpha}{\alpha}} \exp{\left( -\left(\frac{v_0}{c_v\voave}\right)^\frac{1}{\alpha} \right)} .
\end{equation}
Obviously, for a linear dependence, $\alpha=1$, the velocity probability density function is an exponential distribution,
\begin{equation}
    \voave P_{v_0}(v_0;1) = \exp{\left( -\frac{v_0}{\voave} \right)} .
\end{equation}
More generally, for $0<\alpha<1$ the velocity distribution is unimodal, and for small $\alpha$ it resembles a narrow normal distribution. In the limit $\alpha\rightarrow0$ it approaches a degenerate distribution as described above. For $\alpha=1/2$,
\begin{equation}
    \voave P_{v_0}(v_0;1/2) = \frac{\pi v_0}{2\voave} \exp{\left( - \frac{\pi v_0^2}{4\voave^2} \right)} ,
\end{equation}
the velocity probability density function is the Rayleigh distribution.

\begin{figure}[tb]
\includegraphics[width=8cm]{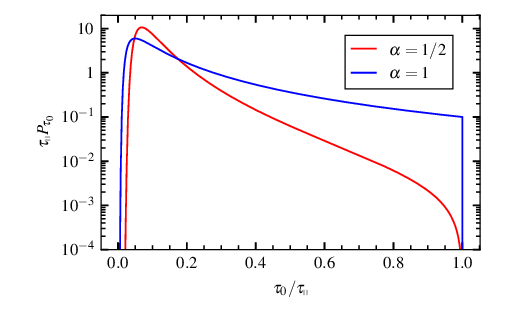}
\caption{Probability density function of pulse durations at the reference position $x=0$ in the case where pulse velocities have a power law dependence on exponentially distributed amplitudes for scaling exponents $\alpha=1/2$ and $\alpha=1$. The normalized linear damping time is $\voave\taup/\ell=10$.}
\label{fig:Pduration}
\end{figure}

Given the velocity probability density function, the distribution of pulse durations at the reference position can be calculated by use of \Eqref{eq.ptau}. For an exponential amplitude distribution and power law amplitude dependence, the pulse duration distribution is non-zero for $0<\tau_0<\taup$ and then given by
\begin{equation}\label{eq.Ptau0}
    P_{\tau_0}(\tau_0;\alpha) = \frac{\ell}{\alpha c_v \voave\tau_0^2} \left( \frac{\ell(\taup-\tau_0)}{c_v\voave\taup\tau_0} \right)^\frac{1-\alpha}{\alpha} \exp\left( -\left( \frac{\ell(\taup-\tau_0)}{c_v\voave\taup\tau_0} \right)^{\frac{1}{\alpha}} \right) .
\end{equation}
For $\alpha>0$, this distribution has a compressed exponential truncation for short pulse durations and a power law scaling as the pulse duration approaches the linear damping time. In the limit $\alpha\rightarrow0$ this becomes a degenerate distribution since all pulses then have the same velocity. For $\alpha=1/2$, the duration distribution is
\begin{equation}
    P_{\tau_0}(\tau_0;1/2) = \frac{\pi\ell^2}{2\voave^2\tau_0^3} \frac{\taup-\tau_0}{\taup} \exp\left( - \frac{\pi\ell^2}{4\voave^2\tau_0^2} \left( \frac{\taup-\tau_0}{\taup} \right)^2 \right) ,
\end{equation}
while for $\alpha=1$ the probability density function is given by
\begin{equation}\label{eq.Ptau0a1}
    P_{\tau_0}(\tau_0;1) = \frac{\ell}{\voave\tau_0^2} \exp\left( - \frac{\ell}{\voave\tau_0}\frac{\taup-\tau_0}{\taup} \right) .
\end{equation}
This gives the average pulse duration in the case $\alpha=1$,
\begin{equation}\label{eq:tauavea=1}
    \langle{\tau_0}\rangle = \frac{\ell}{\voave}\exp\left( \frac{\ell}{\voave\taup} \right) \Gamma\left(0, \frac{\ell}{\voave\taup} \right) ,
\end{equation}
where $\Gamma(\alpha,x)$ denotes the upper incomplete Gamma function defined by
\begin{equation}
    \Gamma(\alpha,x) = \int_x^\infty \text{d}\zeta\,\zeta^{\alpha-1}\exp{(-\zeta)} .
\end{equation}
The pulse duration distribution is presented in \Figref{fig:Pduration} for $\alpha=1/2$ and $\alpha=1$ and a normalized linear damping time $\voave\taup/\ell=10$. The pulse duration is upper limited by the linear damping time. The distribution has an exponential truncation for short durations and is therefore unimodal. In the absence of linear damping, the average pulse duration is finite only for $\alpha<1$.\cite{korzeniowska_longrange_2024}

\begin{figure}[tb]
\includegraphics[width=8cm]{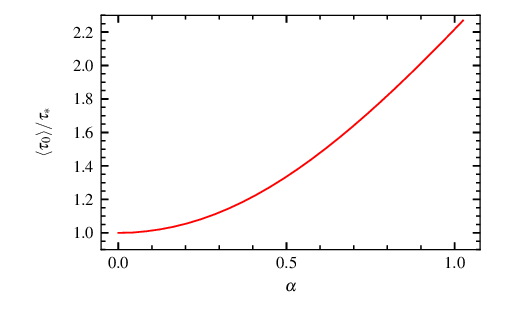}
\caption{Normalized average pulse duration as function of the scaling exponent $\alpha$. The normalized linear damping time is $\voave\taup/\ell=10$.}
\label{fig:taud_alpha}
\end{figure}

From the probability distribution in \Eqref{eq.Ptau0} we can numerically calculate the average pulse duration at the reference position. For $\alpha=0$ this is given by \Eqref{eq.tau} and for $\alpha=1$ by \Eqref{eq:tauavea=1}. The average pulse duration as function of the scaling exponent $\alpha$ is presented in \Figref{fig:taud_alpha}. When the pulse velocities are randomly distributed, the average duration is dominated by the slow pulses and therefore increases with the scaling exponent. For $\alpha=1$, the average duration is more than twice the value in the reference case where all pulses have the same velocity. This increases the degree of pulse overlap at the reference position.

\begin{figure}
\centering
\includegraphics[width=8cm]{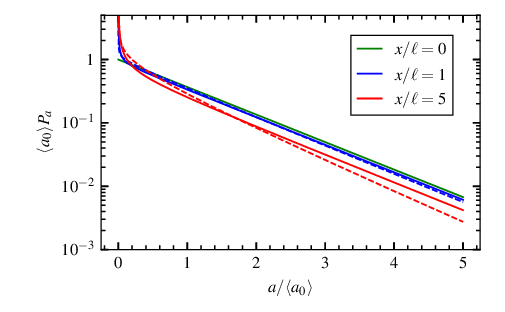}
\caption{Probability density function of the pulse amplitudes at various radial positions for $\alpha=1$ (full lines) and $\alpha=1/2$ (dashed lines). The normalized linear damping time is $\voave\taup/\ell=10$.}
\label{fig:Paxi}
\end{figure}

From \Eqref{eq.axiexp} we can obtain the probability density function for the pulse amplitudes at any radial position. For a general amplitude distribution $P_{a_0}$ at the reference position, the distribution at any position $\xi$ is given by
\begin{equation}\label{eq.Paxi}
         P_{\axi}(\axi) = \frac{\aoave\left(\frac{\alpha\xi}{c_v\voave\taup \Lambda\left(\frac{\alpha \xi \aoave^\alpha}{c_v\voave\taup a^\alpha}\right)}\right)^\frac{1}{\alpha}}{a\left(\Lambda\left(\frac{\alpha \xi \aoave^\alpha}{c_v\voave\taup a^\alpha}\right)+1\right)}\,P_{a_0}\left( \aoave \left(\frac{\alpha\xi}{c_v\voave\taup \Lambda\left(\frac{\alpha \xi \aoave^\alpha}{c_v\voave\taup a^\alpha}\right)}\right)^\frac{1}{\alpha} \right) ,
\end{equation}
where $\Lambda(x)$ denotes the product logarithm function, which returns the principal solution for $\zeta$ in $x=\zeta\exp(\zeta)$ where $\Lambda(x)$ is real for $x>-\exp{(-1)}$. Here we have suppressed the $\xi$ subscript on the amplitude $a$ for simplicity of notation. In the case $\alpha=0$, all pulses have the same velocity and the amplitude distribution is the same at all radial positions with the mean value given by \Eqref{eq.aavex}. The probability density function for $P_\Phi$ is presented in \Figref{fig:Paxi} at various radial positions for an exponential amplitude distribution at the reference position and for $\alpha=1/2$ and $\alpha=1$. The linear damping leads to an abundance of small-amplitude pulses for the downstream process.\cite{losada_stochastic_2024}

\begin{figure}[tb]
\centering
\includegraphics[width=8cm]{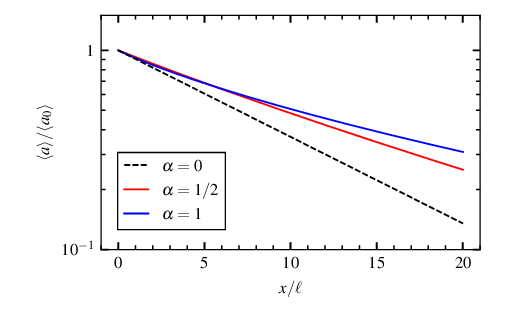}
\caption{Radial variation of the average pulse amplitude for pulse velocities given by a power law dependence on initial amplitudes and an exponential amplitude distribution at the reference position $x=0$ for various scaling exponents $\alpha$. The normalized linear damping time is $\voave\taup/\ell=10$.}
\label{fig:aveaprof}
\end{figure}

From the amplitude distribution in \Eqref{eq.Paxi} we can numerically calculate the radial variation of the average pulse amplitude. This is presented in \Figref{fig:aveaprof} for various values of the scaling exponent in the case $\voave\taup/\ell=10$. The dashed line is the exponential profile in the limit $\alpha\rightarrow0$, given by \Eqref{eq.aavex}. The average amplitude profile becomes flatter for stronger correlation between pulse amplitudes and velocities, which obviously increases the mean value of the process.

\begin{figure*}[tb]
\centering
\includegraphics[width=16cm]{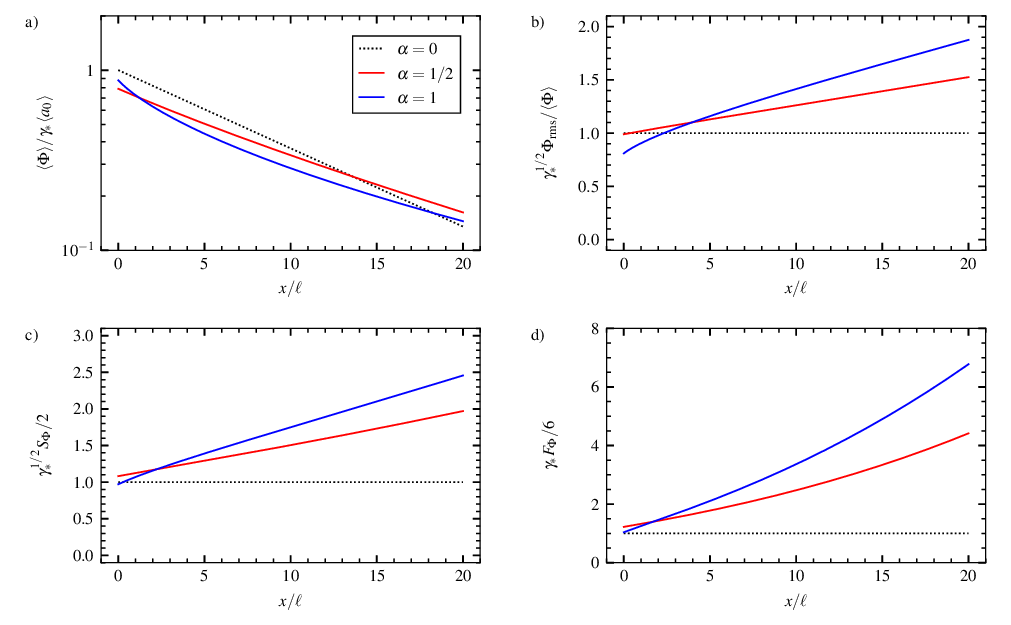}
\caption{Radial variation of the lowest order moments for pulse velocities given by a power law dependence on initial amplitudes and an exponential amplitude distribution at the reference position $x=0$ for various scaling exponents $\alpha$. Plot panels show a) mean value of the process, b) relative fluctuation level, c) skewness moment and d) flatness moment. The dotted lines give profiles for the reference case where all pulses have the same velocity. For all radial profiles, the normalized linear damping time is $\voave\taup/\ell=10$.}
\label{fig.prof_timeindep_Paexp}
\end{figure*}

The radial variation of the lowest order moments of the process in the case of time-independent pulse velocities with a power law dependence on amplitudes is presented in \Figref{fig.prof_timeindep_Paexp} for various values of the scaling exponent $\alpha$ and an exponential amplitude distribution at the reference position. All of these profiles are normalized by their values at the reference position for the reference case of a degenerate distribution of pulse velocities discussed in \Secref{sec.fpp}. The dashed lines show the radial profiles for this reference case. The mean value at $x=0$ is slightly lower for $\alpha=1/2$ and $\alpha=1$ than for the reference case. This is due to correlations between the pulse amplitudes and velocities at the reference positions for $\alpha>0$.\cite{losada_stochastic_2024} The radial variation of the mean value is close to exponential with a comparable e-folding length as for the reference case where all pulses have the same velocity. However, the relative fluctuation level as well as the skewness and flatness moments increase radially outwards and are considerably higher than for the reference case discussed in \Secref{sec.fpp}. This is because the downstream process becomes dominated by the large-amplitude pulses which are subject to less linear damping. In the following section it will be shown that time-dependent pulse velocities lead to an additional mechanism for an increased fluctuation level and higher intermittency of the fluctuations.

\section{Time-dependent velocities}\label{sec.t-dv}

Consider now the case with time-dependent pulse velocities which are given by a power law dependence on the instantaneous pulse amplitude,
\begin{equation}\label{eq.velocity}
    \frac{V(t)}{\voave} = c_v \left( \frac{A(t)}{\aoave} \right)^\alpha ,
\end{equation}
where $\voave$ is the average velocity at the reference position and the time-dependent pulse amplitude is given by \Eqref{eq.amplitude}. The time dependence of the velocity is therefore due to the exponential decrease of the pulse amplitude. Note that capital letters $A(t)$ and $V(t)$ are used to denote the time-dependence of the pulse amplitudes and velocities, while lower case letters $a_{0}=A(0)$ and $v_{0}=V(0)$ denote the initial random variables at the reference position $x=0$. At this position, their relation is thus given by \Eqref{eq.v0a0powerlaw}.

In the limit $\alpha\rightarrow0$, the pulse velocity becomes independent of the amplitude and has a degenerate distribution, that is, all pulses move radially outwards with the same, time-independent velocity. The model then reduces to the reference case, which is the filtered Poisson process discussed in \Secref{sec.fpp}. When $\alpha=1$, a linear equation describes the relationship between $A$ and $V$, and the velocity distribution is the same as the amplitude distribution at all radial positions. In the following, the implications of the time-dependent velocities for the statistical properties of the process will be investigated in detail.

\subsection{Stagnation of pulses}\label{sec.stagnation}

With the power law velocity dependence on the instantaneous pulse amplitude, the radial position of the pulse given by \Eqref{eq.Xk} integrates to
\begin{equation}\label{eq.radialpos}
    X(t) = \Xmax \left[ 1 - \exp{\left( - \frac{\alpha t}{\taup} \right)} \right] ,
\end{equation}
where
\begin{equation}\label{eq.Xmax}
    \Xmax = \frac{c_v \voave \taup}{\alpha} \left( \frac{a_0}{\ave{a_0}} \right) ^\alpha .
\end{equation}
In the limit $t/\taup\rightarrow\infty$, the pulse position approaches the maximum value $\Xmax$. This implies that the pulse stagnates at position $\Xmax$ with its amplitude diminishing exponentially in time as it approaches this position. With use of \Eqref{eq.v0a0powerlaw}, the maximum position can alternatively be written as $\Xmax=v_0\taup/\alpha$. Relative to its size, the distance the pulse travels can thus be written as $v_{0}\taup/\alpha\ell$, that is, the ratio of the linear damping time and the radial transit time based on the initial velocity. Large-amplitude pulses have short radial transit times and undergo less amplitude reduction due to linear damping. Consequently, fast and large-amplitude pulses move further before they stagnate. The maximum pulse position is inversely proportional to the scaling exponent $\alpha$. Accordingly, the stagnation point diverges in the limit $\alpha\rightarrow0$ since the pulse velocity then becomes independent of the amplitude and has no time-dependence.

\begin{figure}[tb]
    \centering
    \includegraphics[width=8cm]{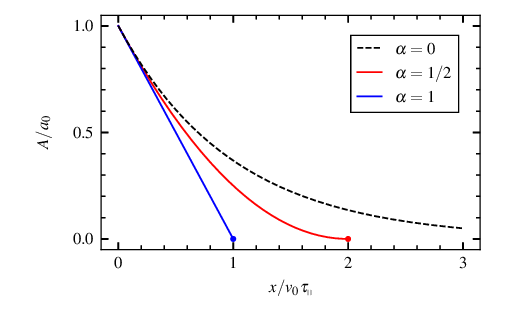}
    \caption{Radial variation of the pulse amplitude for various values of the scaling exponent $\alpha$. The dashed line is the exponential variation in the case of independent amplitudes and velocities ($\alpha=0$). The filled circles at $A=0$ indicates the stagnation points for time-dependent pulses. The linear damping time is $v_0\taup/\ell=10$.}
    \label{fig:maxa(x)}
\end{figure}

The time $s_\xi-s_0$ it takes for a pulse to move from the reference position at time $s_0$ to any radial position $\xi<\Xmax$ is given by the relation
\begin{equation}
    \xi = \int_{s_0}^{s_\xi} \text{d}t\,V(t) = \Xmax\left[ 1-\exp\left( -\frac{\alpha (s_\xi-s_0)}{\taup} \right) \right] .
\end{equation}
From this, it follows that the pulse arrival time at the position $\xi$,
\begin{equation}\label{eq.sxi}
    s_\xi = s_0 - \frac{\taup}{\alpha}\ln{\left(1-\frac{\xi}{\Xmax}\right)} ,
\end{equation}
diverges logarithmically as the stagnation point is approached. Using this relation between the pulse position and arrival time, the pulse amplitude given by \Eqref{eq.amplitude} can be written as a function of the radial position,
\begin{equation}\label{eq.max.A(xi)}
    \frac{A_{\xi}}{a_{0}} = \frac{A(s_\xi-s_0)}{a_{0}} = \left( 1-\frac{\xi}{\Xmax} \right)^{\frac{1}{\alpha}} .
\end{equation}
The radial amplitude variation is presented in \Figref{fig:maxa(x)} for various values of the scaling exponent $\alpha$. For time-independent pulses, $\alpha=0$, we have $A_{\xi}/a_{0}=\exp{(-\xi/v_{0}\taup)}$, the familiar exponential amplitude profile for time-independent velocities given by \Eqref{eq.axiexp}. For $\alpha=1$ the amplitude decreases linearly with radius.

The radial variation of the pulse velocity similarly follows from \Eqref{eq.velocity},
\begin{equation}\label{eq.Vxi}
    \frac{V_\xi}{v_0} = \frac{V(s_\xi-s_0)}{v_0} = 1-\frac{\xi}{\Xmax} .
\end{equation}
This decreases linearly with radius for all values of the scaling exponent $\alpha$. Again, in the limit $\alpha\rightarrow0$, $\Xmax$ diverges and the pulse velocity does not change with radius since the velocity is time-dependent.

\begin{figure}[tb]
\centering
\includegraphics[width=8cm]{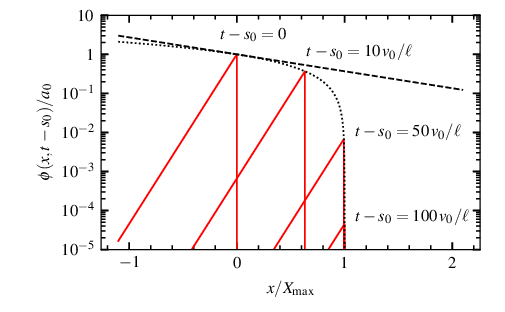}
\caption{Radial variation of a one-sided exponential pulse at various times for a linear relation between pulse velocity and amplitude ($\alpha=1$). The dotted line shows the radial variation of the peak pulse amplitude and the dashed line is the exponential decrease for time-independent pulses ($\alpha=0$). The normalized linear damping time is $v_0 \taup /\ell = 10$.}
\label{fig.stagnation}
\end{figure}

The radial variation of an exponential pulse with linear velocity dependence on the exponentially decreasing pulse amplitude is presented in \Figref{fig.stagnation} at various times. As the stagnation point is approached, the pulse amplitude diminishes due to linear damping. The dotted line corresponds to the peak amplitude variation with radial position as given by \Eqref{eq.max.A(xi)} and the dashed line show the exponential amplitude variation in the case of a time-independent pulse velocity. Before analyzing the implications of such pulse stagnation on the rate of pulses, we first discuss the temporal scales involved in the pulse evolution.

\subsection{Pulse durations}

The super-position of exponential pulses at any position $\xi$ can be written as
\begin{equation}
    \Phi_{K_\xi}(\xi,t) = \sum_{k=1}^{K_\xi} a_{0k}\exp\left(-\frac{t-s_{0k}}{\taup}\right) \varphi\left(\frac{\xi-X_k(t-s_{0k})}{\ell}\right) ,
\end{equation}
where the pulse position $X$ is given by \Eqref{eq.radialpos} and $K_\xi$ is the number of pulses reaching this position. By introducing the pulse arrival times $s_\xi$ at this position, given by \Eqref{eq.sxi}, and the radial variation of the pulse amplitude $A_\xi$, given by \Eqref{eq.max.A(xi)}, each pulse can be written as
\begin{equation}\label{eq.phixi}
    \phi(\xi,t-s_\xi) = A_{\xi}\exp\left( -\frac{t-s_\xi}{\taup} \right) \varphi\left( \frac{\xi}{\ell} + \frac{\Xmax}{\ell}\left[ \left(1-\frac{\xi}{\Xmax}\right)\exp\left( -\frac{\alpha(t-s_\xi)}{\taup} \right)-1 \right] \right) .   
\end{equation}
Unlike the case with time-independent velocities in \Eqref{eq.Phixi}, the explicit radial dependence does not disappear by introducing the variables $s_\xi$ and $A_\xi$ at the pulse position. However, the relevant asymptotic limits can readily be obtained.

For short times compared to the linear damping time, $0<t-s_\xi\ll\taup$, the exponential inside the argument of the pulse function in \Eqref{eq.phixi} can be expanded to give the pulse $\phi$ on the familiar form $A_\xi\exp(-(t-s_\xi)/\tau_\xi)$,
where the average pulse duration has radial dependence,
\begin{equation}\label{eq.tauxi}
    \tau_\xi = \frac{\taup\ell}{V_\xi\taup+\ell} ,
\end{equation}
with $V_\xi$ given by \Eqref{eq.Vxi}. This is the same harmonic mean of the radial transit time and linear damping time as for the time-independent case, but with the local pulse velocity given by \Eqref{eq.Vxi}. Since the pulse velocity decreases radially outwards, the duration becomes longer. At the reference position $\xi=0$, the initial pulse evolution is the same as for the case of time-independent velocities discussed in \Secref{sec.tiv}.

For long times compared to the linear damping time, $t-s_\xi\gg\taup$, the pulse has nearly reached its stagnation point and the temporal evolution at any position $\xi$ is approximated by
\begin{equation}
    \phi(\xi,t-s_\xi) = a_0\left( 1+\frac{\xi}{\Xmax} \right)^{\frac{1}{\alpha}} \exp\left( -\frac{\Xmax}{\ell}\left( 1-\frac{\xi}{\Xmax}\right) \right) \exp\left( - \frac{t-s_\xi}{\taup} \right) .
\end{equation}
This reveals an exponential decrease in time given by the linear damping time. This is because the pulse is nearly stagnant as it approaches $\Xmax$, and the radial motion does not contribute to the amplitude decay in this limit, as illustrated in \Figref{fig.stagnation}. At the reference position $\xi=0$, the long time pulse response is
\begin{equation}
    \phi(0,t-s_0) = a_0\exp\left( - \frac{\Xmax}{\ell}\right) \exp\left( - \frac{t-s_0}{\taup} \right) .
\end{equation}
The pulse amplitude is then reduced by an exponential factor determined by the ratio of the distance the pulse travels and its size. Close to the stagnation point, $\xi\rightarrow\Xmax$, the pulse function approaches $A_\xi\exp(-(t-s_\xi)/\taup)$, with an amplitude reduction determined by \Eqref{eq.max.A(xi)}.

\begin{figure}[tb]
\centering
\includegraphics[width=8cm]{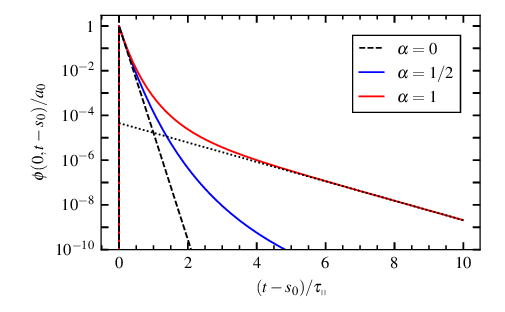}
\caption{Temporal pulse evolution at the reference position $x=0$ for various values of the scaling exponent $\alpha$ and normalized linear damping time $v_0\taup/\ell=10$. The dashed line is the exponential pulse function for the reference case $\alpha=0$, while the dotted line show an exponential decay with the linear damping time.}
\label{fig.duration}
\end{figure}

The temporal pulse evolution at the reference position as given by \Eqref{eq.phixi} is presented in \Figref{fig.duration} for various values of the scaling exponent $\alpha$ and normalized linear damping time $v_0\taup/\ell=10$. The initial exponential decrease is given by the pulse duration in \Eqref{eq.tau}. The asymptotic long time behavior is an exponential decrease at a rate given by the linear damping time $\taup$. For large-amplitude pulses, $v_0\taup/\ell\gg1$, the slow asymptotic behavior does not set in until the pulse amplitude has been reduced substantially due to linear damping. For small-amplitude pulses, the linear damping time is comparable to or shorter than the radial transit time, and the temporal scales is determined by the linear damping time. Since $\voave\taup/\ell$ is typically large, most pulses recorded at any radial position are therefore well described by a one-sided exponential function.

A random distribution of pulse velocities also leads to a distribution of pulse durations. From \Eqref{eq.tauxi} follows the pulse duration distribution at the radial position $\xi$, which for $0<\tau_\xi<\taup$ is given by
\begin{multline}\label{eq.Ptauxi}
    P_{\tauxi}(\tauxi;\alpha) = \frac{\ell}{\alpha c_v \voave\tauxi^2} \left( \frac{\ell(\taup-\tauxi)}{c_v\voave\taup\tauxi} + \frac{\alpha\xi}{c_v\voave\taup}\right)^\frac{1-\alpha}{\alpha} \\ \exp\left( -\left( \frac{\ell(\taup-\tauxi)}{c_v\voave\taup\tauxi} + \frac{\alpha\xi}{c_v\voave\taup}\right)^{\frac{1}{\alpha}} +\left(\frac{\alpha\xi}{c_v\voave\taup}\right)^{\frac{1}{\alpha}} \right) .  
\end{multline}
At the reference position $\xi=0$ this becomes the same distribution as for the case of time-independent pulse velocities given by \Eqref{eq.Ptau0} and presented in \Figref{fig:Pduration}. In the absence of linear damping, this distribution has a power law scaling $1/\tau^{1+1/\alpha}$ for long durations and an exponential truncation for short durations.

\begin{figure}[tb]
\centering
\includegraphics[width=8cm]{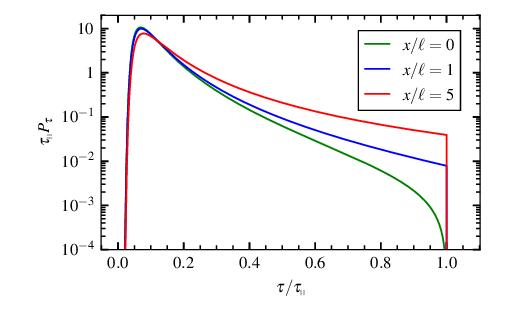}
\caption{Probability density function of the pulse durations at various radial positions for $\alpha=1/2$. The normalized linear damping time is $\voave\taup/\ell=10$.}
\label{fig.Ptauxi}
\end{figure}

For $\alpha=1/2$ the duration distribution is given by
\begin{multline}\label{eq:Ptauxia=1/2}
    P_{\tauxi}(\tauxi;1/2) = \frac{\pi\ell}{\voave\tauxi^2} \left( \frac{\ell(\taup-\tauxi)}{2\voave\taup\tauxi} + \frac{\xi}{4\voave\taup}\right) \\ \exp\left( - \pi\left( \frac{\ell(\taup-\tauxi)}{2\voave\taup\tauxi} + \frac{\xi}{4\voave\taup}\right)^2 + \pi\left(\frac{\xi}{4\voave\taup}\right)^2 \right) .  
\end{multline}
The probability density function for $\alpha=1/2$ is presented in \Figref{fig.Ptauxi} for various radial positions. These distributions are unimodal, have an exponential truncation for short durations and a hard truncation at the linear damping time.

\begin{figure}[tb]
\centering
\includegraphics[width=8cm]{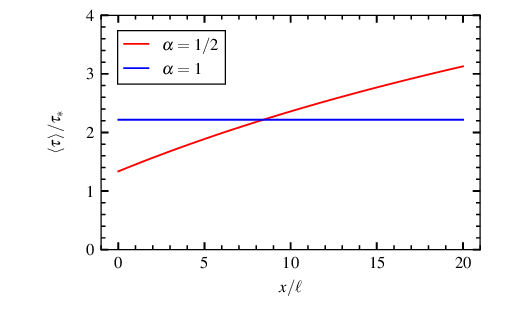}
\caption{Radial variation of the average pulse duration for various values of the scaling exponent $\alpha$. The normalized linear damping time is $\voave\taup/\ell=10$.}
\label{fig.avetauxi}
\end{figure}

For $\alpha=1$, the $\xi$ dependence in \Eqref{eq.Ptauxi} cancels and the distribution is thus the same for all radial positions and for $0<\tau<\taup$ given by
\begin{equation}\label{eq.Ptauxia0}
    P_{\tauxi}(\tauxi;1) = \frac{\ell}{\voave\tauxi^2} \exp\left( - \frac{\ell}{\voave\tauxi}\frac{\taup-\tauxi}{\taup} \right) .  
\end{equation}
This is the same as in \Eqref{eq.Ptau0a1} for the reference position. As shown in \Figref{fig:Pduration}, this distribution has an exponential truncation for short durations and a power law scaling $1/\tau^2$ as the pulse duration approaches the linear damping time. 
The average pulse duration is presented in \Figref{fig.avetauxi} for various values of the scaling exponent $\alpha$ and a linear damping time $\voave\taup/\ell=10$. For $\alpha=1/2$, the average pulse duration increases radially outwards due to an abundance of slow pulses. However, for $\alpha=1$ this is cancelled by stagnation of small-amplitude pulses and the average duration is independent of the radial position, as shown from the probability distribution above, and i
given by \Eqref{eq:tauavea=1}. In the limit $\alpha\rightarrow0$, all pulses have the same velocity and the duration is the same for all of them.

\subsection{Average amplitudes and velocities}

From \Eqref{eq.max.A(xi)} we can obtain the probability density function for the pulse amplitudes at any given radial position. For an exponential amplitude distribution at the reference position, the amplitude distribution at any position $\xi$ is for $\Axi>0$ given by
\begin{equation}\label{eq.PAxi}
    \aoave P_{\Axi} (\Axi;\alpha) = \left(1+\frac{\alpha\xi}{c_v\voave\taup}\frac{\aoave^\alpha}{\Axi^\alpha}\right)^{\frac{1-\alpha}{\alpha}} \exp\left(- \left(\frac{\Axi^\alpha}{\aoave^\alpha}+\frac{\alpha\xi}{c_v\voave\taup}\right)^{\frac{1}{\alpha}} + \left(\frac{\alpha\xi}{c_v\voave\taup}\right)^{\frac{1}{\alpha}} \right) .
\end{equation}
Here and in the following, we suppress the $\xi$ subscript on the pulse amplitude $A$ and velocity $V$ for simplicity of notation. Equation~\eqref{eq.PAxi} is a generalization of \Eqref{eq.Paxi} for the case of time-dependent pulse velocities. For $\xi=0$ this is just the exponential distribution at the reference position given by \Eqref{eq.pdfaoexp}. As discussed previously, the limit $\alpha\rightarrow0$ corresponds to time-independent velocities, and the amplitude distribution is then exponential at all radial positions with the mean value given by \Eqref{eq.aavex},
\begin{equation}
    \aoave\exp(-\xi/\voave\taup)\,P_\Axi(\Axi;0) = \exp\left( - \frac{\Axi}{\aoave\exp(-\xi/\voave\taup)} \right) .
\end{equation}
From the distribution in \Eqref{eq.PAxi} it furthermore follows that for $\alpha=1$ the amplitude distribution is exponential and for $\Axi>0$ given by
\begin{equation}\label{eq:PAalpha1}
    \aoave P_{\Axi}(\Axi;1) = \exp\left( - \frac{\Axi}{\aoave} \right) .
\end{equation}
Thus, in this case the average amplitude is radially constant and given by the value at the reference position, $\aoave$. This is clearly due to the stagnation of small-amplitude pulses.

\begin{figure}[tb]
\centering
\includegraphics[width=8cm]{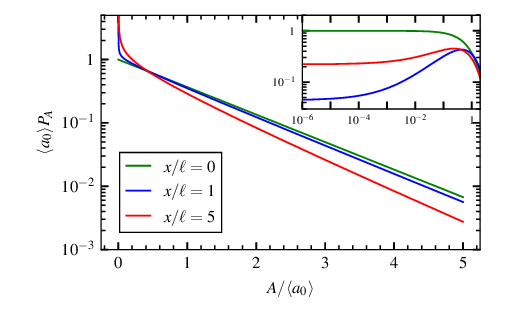}
\caption{Probability density function of the pulse amplitudes at various radial positions for $\alpha=1/2$. The inset shows the compensated distributions $A^{1/2}P_A$ with double logarithmic scales. The normalized linear damping time is $\voave\taup/\ell=10$.}
\label{fig.PAxi}
\end{figure}

More generally, the distribution given by \Eqref{eq.PAxi} has an exponential tail for large amplitudes and a power law scaling $P_A\sim A^{\alpha-1}$ for small amplitudes. For $\alpha=1/2$ the probability distribution is given by
\begin{equation}\label{eq.P_A1/2}
    \aoave P_\Axi(\Axi;1/2) = \left( 1 + \frac{\pi^{1/2}\xi}{4\voave\taup}\frac{\aoave^{1/2}}{\Axi^{1/2}} \right) \exp\left( - \left( \frac{\Axi^{1/2}}{\aoave^{1/2}} + \frac{\pi^{1/2}\xi}{4\voave\taup} \right)^2 + \left( \frac{\pi^{1/2}\xi}{4\voave\taup} \right)^2 \right) .
\end{equation}
This distribution is presented in \Figref{fig.PAxi} for various radial positions, clearly displaying an exponential tail for large amplitudes at all positions. The inset in \Figref{fig.PAxi} shows the compensated distribution $A^{1/2}P_A(A;1/2)$ with double logarithmic scaling, revealing the power law scaling $P_A(A;1/2)\sim A^{-1/2}$ for low amplitudes in \Eqref{eq.P_A1/2}.

\begin{figure}[tb]
\centering
\includegraphics[width=8cm]{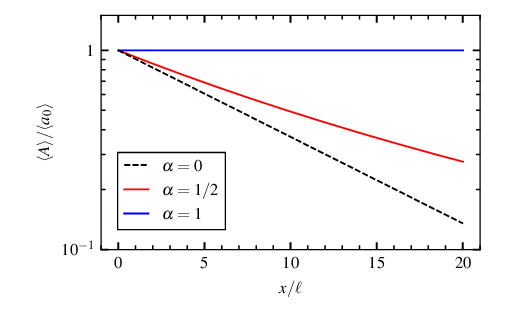}
\caption{Radial variation of the average pulse amplitude for various values of the scaling exponent $\alpha$. The normalized linear damping time is $\voave\taup/\ell=10$.}
\label{fig:<a>(x)}
\end{figure}

The radial variation of the average amplitude is given by
\begin{equation}\label{eq:Aavea=1/2}
    \frac{\langle{\Axi}\rangle}{\aoave} = 1-\frac{\pi\xi\exp\left(\left(\frac{\pi^{1/2}\xi}{4\voave\taup}\right)^2\right)\erfc{\left(\frac{\pi^{1/2}\xi}{4\voave\taup}\right)}}{4\voave\taup} .
\end{equation}
This is presented in \Figref{fig:<a>(x)} for various values of the scaling exponent $\alpha$. The dashed line is exponential mean amplitude profile in the case of time-independent pulse velocities in the limit $\alpha\rightarrow0$. 
Stagnation of small amplitude pulses for $\alpha>0$ implies an average amplitude that decreases weaker than exponential.  For $\alpha=1$, pulse stagnation balances the linear damping and results in a radially constant average amplitude.

Similarly, from \Eqref{eq.Vxi} follows the velocity distribution at the radial position $\xi$, which for $V_\xi>0$ is given by
\begin{equation}\label{eq.PVxi}
        \voave P_{\Vxi}(\Vxi;\alpha) = 
        \frac{1}{\alpha c_v}\left(\frac{\Vxi}{c_v\voave}+\frac{\alpha\xi}{c_v\voave\taup}\right)^{\frac{1-\alpha}{\alpha}}\exp\left( - \left(\frac{\Vxi}{c_v\voave}+\frac{\alpha\xi}{c_v\voave\taup}\right)^{\frac{1}{\alpha}} + \left(\frac{\alpha\xi}{c_v\voave\taup}\right)^{\frac{1}{\alpha}} \right) .
\end{equation}
In the limit $\alpha\rightarrow0$ this becomes a narrow distribution centered at $\voave$, corresponding to the reference case where all pulses have the same velocity. For $\alpha=1$, the pulse velocities are proportional to the amplitudes and this becomes an exponential distribution, which for $\Vxi>0$ is given by
\begin{equation}
    \voave P_{\Vxi}(\Vxi;1) = \exp\left( - \frac{\Vxi}{\voave} \right) .
\end{equation}
Thus, also in this case the average velocity is constant as function of radius, $\Vave=\voave$, similar to the average pulse amplitude described by \Eqref{eq:PAalpha1}.

\begin{figure}[tb]
\centering
\includegraphics[width=8cm]{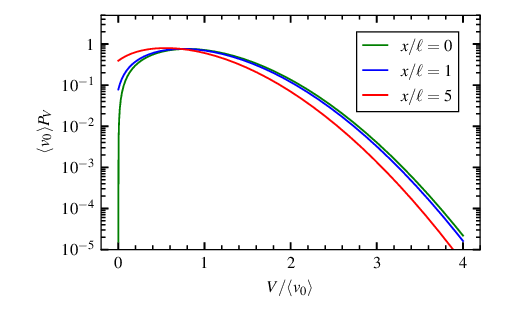}
\caption{Probability density function of the pulse velocities for $\alpha=1/2$ at various radial positions. The normalized linear damping time is $\voave\taup/\ell=10$.}
\label{fig.PVxi}
\end{figure}

For the case $\alpha=1/2$, the velocity probability density function is given by
\begin{equation}
        \voave P_{\Vxi}(\Vxi;1/2) = 
        \pi \left(\frac{\Vxi}{2\voave}+\frac{\xi}{4\voave\taup}\right) \exp\left( - \pi\left(\frac{\Vxi}{2\voave}+\frac{\xi}{4\voave\taup}\right)^2 + \pi\left(\frac{\xi}{4\voave\taup}\right)^2 \right) .
\end{equation}
The distribution is presented in \Figref{fig.PVxi} for various radial positions in the case $\voave\taup/\ell=10$. At the reference position $\xi=0$ this probability density is the same as in the case of time-independent velocities, which is a Rayleigh distribution as discussed in \Secref{sec:corrav}.

\begin{figure}[tb]
\centering
\includegraphics[width=8cm]{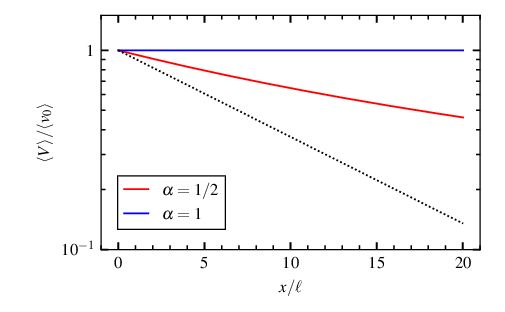}
\caption{Radial variation of the average pulse velocity for various values of the scaling exponent $\alpha$. The dotted line is the exponential decay $\exp(-x/\voave\taup)$. The normalized linear damping time is $\voave\taup/\ell=10$.}
\label{fig.aveVxi}
\end{figure}

From the velocity distribution function in \Eqref{eq.PVxi} follows the radial variation of the average velocity for general $\alpha$,
\begin{equation}\label{eq:Vave}
    \frac{\langle{\Vxi}\rangle}{\voave} = \alpha c_v \exp \left(\left[\frac{\alpha\xi}{c_v\voave\taup}\right]^{\frac{1}{\alpha}}\right)\Gamma\left(\alpha, \left[\frac{\alpha\xi}{c_v\voave\taup}\right]^{\frac{1}{\alpha}}\right) .
\end{equation}
For $\alpha=1/2$ the radial profile of the average velocity is given by
\begin{equation}\label{eq:Vavea=1/2}
    \frac{\Vave}{\voave}=\exp\left(\left(\frac{\pi^{1/2}\xi}{4\voave\taup}\right)^2\right)\erfc{\left(\frac{\pi^{1/2}\xi}{4\voave\taup}\right)} .
\end{equation}
The radial variation of the average velocity is presented in \Figref{fig.aveVxi} for various values of the scaling exponent $\alpha$ in the case $\voave\taup/\ell=10$. Similar to the average amplitude profile, stagnation of small-amplitude pulses for $\alpha>0$ implies an average velocity that decreases weaker than exponential and becomes radially constant for $\alpha=1$.

\subsection{Pulse waiting times}

\begin{figure}[tb]
\centering
\includegraphics[width=8cm]{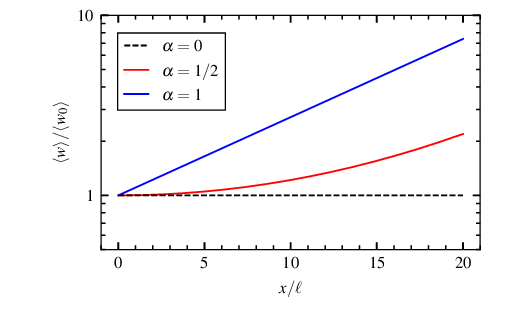}
\caption{Radial variation of the average pulse waiting time for various values of the scaling exponent $\alpha$. The normalized linear damping time is $\voave\taup/\ell=10$.}
\label{fig:prof_tauw_exp}
\end{figure}

Pulse stagnation implies that the rate of pulses will decrease with radial position in the process, corresponding to a radial increase in the average pulse waiting time. This is straight forward to quantify by calculating the number of pulses that leave at $x=0$ during a time interval T and eventually reach the position $\xi$,
\begin{equation}
    K_\xi = \sum_{k=1}^{K_0} \Theta\left( X_{\text{max}\,k}-\xi \right) .
\end{equation}
Here $K_0$ is the total number of pulses at the reference position and $\Theta$ denotes the unit step function. Since the pulses are uncorrelated, the sum is equal to $K_0\Theta(\Xmax-\xi)$. Integrating over the amplitude distribution and the number of pulses then gives
\begin{equation}\label{eq.Kxiave}
    \Kxiave = \Koave\int_0^\infty \text{d}a_0\,P_{a_0}(a_0)\Theta\left( \frac{c_v\voave\taup}{\alpha}\left(\frac{a_0}{\aoave}\right)^\alpha - \xi \right) .
\end{equation}
The step function gives a contribution only when its argument is positive, implying a lower limit on the pulse amplitude given by
\begin{equation}
    \aomin = \aoave \left( \frac{\alpha\xi}{c_v\voave\taup} \right)^{\frac{1}{\alpha}} .
\end{equation}
In the stationary state, the average pulse waiting time at the radial position $\xi$ can be written as $\tauwxi=T/\Kxiave$. Performing the averaging over an exponential distribution of pulse amplitudes in \Eqref{eq.Kxiave} thus gives the radial variation of the average waiting time,
\begin{equation}\label{eq.tauwxi}
    \tauwxi = \tauw\exp\left( \left[ \frac{\alpha\xi}{c_v\voave\taup} \right]^{\frac{1}{\alpha}}\right) ,
\end{equation}
which is a stretched exponential function. The radial variation of the average pulse waiting time is presented in \Figref{fig:prof_tauw_exp} for various values of the scaling exponent $\alpha$. For $\alpha = 1$, the average waiting time has an exponential increase with radius with e-folding length $\voave\taup$, the same as for the mean value of the process in the case of a degenerate distribution of pulse velocities. For $\alpha=1/2$, the radial variation is significantly weaker with less pulse stagnation. In the limit $\alpha\rightarrow0$ there is no radial variation since the pulse velocities are then time-independent. Clearly, pulse stagnation and radial variation of the intensity of pulses will influence the moments of the process.

\section{Moments and radial variation}\label{sec.profiles}

In this section, we derive a general expression for the cumulants of the process in the case of time-dependent pulse velocities given by \Eqref{eq.velocity}. In general, these have to be calculated numerically. However, in the case of a linear relationship between pulse amplitudes and velocities, closed form analytical expressions can be obtained that provide significant insight to the statistical properties of the process.

\subsection{Mean value of the process}

With time-dependent pulse velocities given by a power law dependence on the instantaneous amplitudes, the solution for the pulse function can be written as
\begin{equation}
    \phi(x,t-s_0) = a_0 \exp \left( -\frac{t-s_0}{\taup}\right) \varphi\left( \frac{x}{\ell} + \frac{\Xmax}{\ell} \left[ \exp{\left( - \frac{\alpha(t-s_0)}{\taup}\right)} - 1 \right] \right) ,
\end{equation}
where $\Xmax$ is given by \Eqref{eq.Xmax}. The mean value of the process can be calculated by averaging the sum of pulses over all random variables. For a stationary process, end effects can be neglected by extending the integration limits for the pulse arrival times to infinity. Changing the integration variable from the uniformly distributed pulse arrivals $s_0$ to $u=\exp(-(t-s_0)/\taup)$ gives the integral of the pulse function,
\begin{equation}\label{eq.avephi3}
    \frac{1}{T}\int_{-\infty}^{\infty} \text{d}s_0\,\phi(x,t-s_0) = \frac{\taup}{T}\int_0^{\infty} \text{d}u\,a_0\varphi\left( \frac{x}{\ell} + \frac{\Xmax}{\ell}(u^\alpha-1) \right) .
\end{equation}
In the case of a one-sided exponential pulse function as defined by \Eqref{eq.phiexp}, the integrand in \Eqref{eq.avephi3} will vanish when the argument of the pulse function is positive, or equivalently when $u>\umax$, with
\begin{equation}\label{eq.umax}
    \umax(x) = \left( 1 - \frac{x}{\Xmax} \right)^{\frac{1}{\alpha}} ,
\end{equation}
where the term on the right hand side is just the same as the amplitude reduction factor in \Eqref{eq.max.A(xi)}. For any given radial position, a sufficiently small value of $u$ means that $\Xmax$ must be sufficiently large. Since $\Xmax$ increases with the pulse amplitude $a_0$, this implies a lower limit on the amplitude. Any pulse will contribute to the mean value at position $x$ only if $\Xmax>x$, thus the minimum amplitude is again given by
\begin{equation}\label{eq.amin}
    \aomin(x) = \aoave\left( \frac{\alpha x}{c_v \voave \taup} \right)^{\frac{1}{\alpha}} .
\end{equation}
Pulses with amplitudes smaller than $\aomin(x)$ will never reach the radial position $x$. With these considerations, the general expression for the radial variation of the mean value of the process is given by
\begin{equation}\label{eq.Phiave}
    \ave{\Phi}(x) = \frac{\taup}{\tauw}\ave{\Theta\left( a_0-\aomin \right) \int_{0}^{\umax} \text{d}u \; a_0 \exp \left( \frac{x}{\ell} + \frac{c_v \voave \taup}{\alpha \ell} \left( \frac{a_0}{\ave{a_0}} \right)^\alpha (u^\alpha -1) \right)} ,
\end{equation}
where the angular brackets denote integration over the amplitude distribution at the reference position. Here it should be noted that the minimum pulse amplitude $\aomin$ depends on the radial coordinate, while $\umax$ depends both on $x$ and the pulse amplitude $a_0$. The radial variation of the mean value will be discussed further in the following two subsections.

\subsection{Cumulants and moments}

Following a similar procedure as for calculating the mean value of the process, we obtain a general expression for the cumulant of any order $n$,
\begin{equation}\label{eq.cumulants.pl}
    \kappa_n(x) = \frac{\taup}{\tauw} \ave{ \Theta\left( a_0-\aomin \right) \int\limits_{0}^{\umax} \rmd u\,a_0^n u^{n-1} \exp\left( \frac{nx}{\ell} + \frac{nc_v \voave \taup}{\alpha \ell} \left( \frac{a_0}{\ave{a_0}} \right)^\alpha \left( u^\alpha - 1 \right) \right)} .
\end{equation}
In general, the cumulants must be computed numerically for any given distribution of pulse amplitudes $a_0$. However, when all pulses have the same size, the amplitudes are exponentially distributed, and the pulse velocities are given by \Eqref{eq.velocity}, the cumulants for the case $\alpha=1$ can be calculated in closed analytically form and are given by 
\begin{equation}\label{eq.kappana1}
    \kappa_{n}(x) = \frac{\ave{a_0}^n(n-1)!}{\tauw} \frac{\taup\ell}{n\ave{v_0}\taup + \ell} \exp\left ( -\frac{x}{\ave{v_0}\taup}\right) .
\end{equation}
This resembles the result for time-independent pulse velocities given in \Eqref{eq.kappan.degenerate}, except the cumulant order $n$ appears in the effective pulse duration $\taup\ell/(n\voave\taup+\ell)$.

From \Eqsref{eq.Phiave} and \eqref{eq.kappana1} it follows that for $\alpha=1$, the mean value of the process for time-dependent velocities is exactly the same as for the time-independent case where all pulses have the same velocity, discussed in \Secref{sec.fpp},
\begin{equation}\label{eq.Phiavetd}
    \Phiave(x) = \frac{\taustar}{\tauw}\,\aoave\exp{\left( -\frac{x}{\voave\taup} \right)} ,
\end{equation}
where the pulse duration $\taustar$ is defined with the average pulse velocity by \Eqref{eq.taustar}. The anticipated reduction of the mean value due to stagnation of pulses is exactly balanced by the distributions of and strong correlation between pulse velocities and amplitudes. As noted previously, fast and large-amplitude pulses will have short radial transit times and contribute significantly to the mean values at all radial positions. In the case of a linear correlation between velocities and amplitudes, this leads to an exponential variation of the mean value of the process, while the average pulse amplitude, velocity and duration are all radially constant. For $\alpha=1/2$, the mean value of the process for time-dependent velocities  can be written in closed form as
\begin{equation}\label{eq:Phiave-a=1/2}
    \ave{\Phi}(x)= \frac{\ave{a_0}}{\tauw}\frac{\pi}{4}\frac{\ell}{\ave{v_0}}\left[ \text{erfc} \left (\frac{\pi^{1/2} x}{4 \ave{v_0}\taup}\right) - \exp \left ( \left(  \frac{2\ave{v_0}\taup}{\pi^{1/2}\ell}\right)^2 + \frac{x}{\ell}\right)\text{erfc} \left (\frac{2\ave{v_0}\taup}{\pi^{1/2}\ell} + \frac{\pi^{1/2} x}{4 \ave{v_0}\taup}\right) \right] ,
\end{equation}
where $\erfc$ denotes the complimentary error function. The radial variation of the mean value of the process is presented in \Figref{fig:prof_exp} for various scaling exponents $\alpha$ and normalized linear damping $\voave\taup/\ell=10$.

Due to the weighting of the pulse velocity with the cumulant order in the effective duration in \Eqref{eq.kappana1}, higher order cumulants are not the same as for the case with a degenerate distribution of pulse velocities. The effective pulse duration will be shorter for higher order cumulants, thereby increasing the relative fluctuation level and the skewness and flatness moments of the process.

\subsection{Radial variation of moments}

\begin{figure*}[tb]
\centering
\includegraphics[width=16cm]{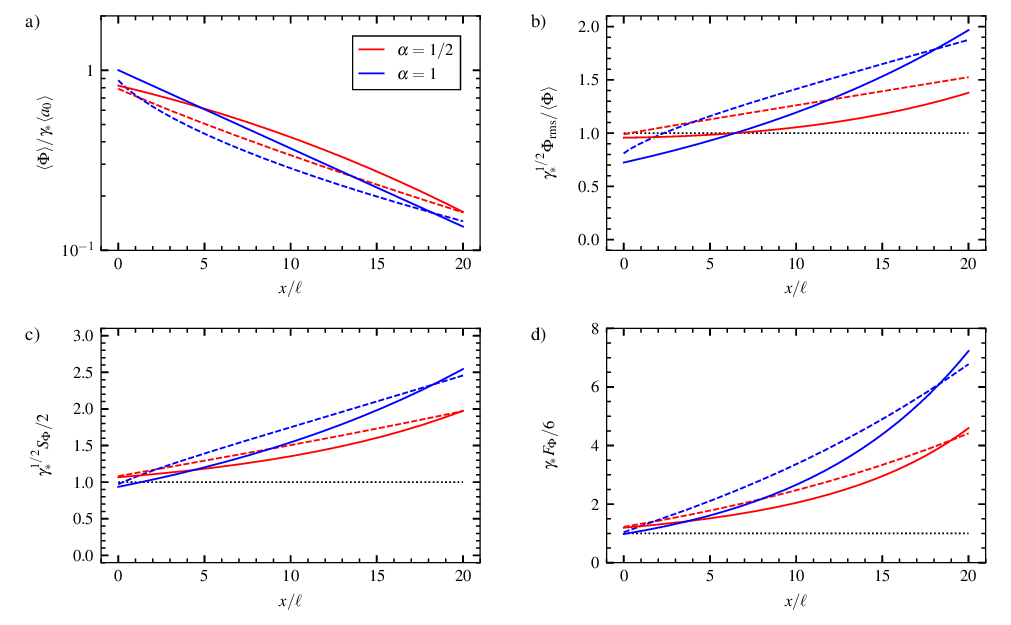}
\caption{Radial variation of the lowest order moments for pulse velocities given by a power law dependence on instantaneous amplitudes and an exponential amplitude distribution at the reference position for various scaling exponents. Plot panels show a) mean value of the process, b) relative fluctuation level, c) skewness moment and d) flatness moment. The dashed lines are the profiles for the case of time-independent pulse velocities, and the dotted lines are the profiles for the reference case where all pulses have the same velocity $(\alpha=0)$. For all radial profiles, the normalized linear damping time is $\voave\taup/\ell=10$.}
\label{fig:prof_exp}
\end{figure*}

The radial variation of the mean value, the relative fluctuation level, and the skewness and flatness moments are presented in \Figref{fig:prof_exp} for an exponential pulse amplitude distribution and time-dependent pulse velocities. The dashed lines give the profiles for the case of time-independent pulse velocities, corresponding to the case $\alpha=0$. As discussed above, for $\alpha=1$ the mean value is an exponential function of radial position with e-folding length $\voave\taup$. As opposed to the reference case where all pulses have the same velocity, the relative fluctuation level as well as the skewness and flatness moments increase radially outwards. These profiles are qualitatively similar to the case of time-independent pulses discussed in \Secref{sec:corrav}. Thus, correlated pulse amplitudes and velocities lead to the same flattened mean profiles as for the reference case but much higher fluctuation amplitudes.

\section{Discussion}\label{sec.disc}

Radial motion of blob-like filament structures at the boundary of magnetically confined plasmas leads to flattened time-average radial profiles in the scrape-off layer with large relative fluctuations. The stochastic modeling presented here and in preceding papers reveals how the radial variation of the lowest order statistical moments depends on the parameters of the blob structures and their correlations. The cross-field transport is modeled as a super-position of uncorrelated pulses that move radially outwards while subject to linear damping due to losses along magnetic field lines. The scalar input parameters for the model are the pulse size $\ell$ and the linear damping time $\taup$, as well as the average pulse amplitude $\aoave$, velocity $\voave$ and waiting time $\woave$ at the reference position, and the exponent $\alpha$ for the power law scaling relationship between pulse velocities and amplitudes. For an exponential pulse function and exponentially distributed pulse amplitudes at the reference position, the model determines the distribution of and correlations between pulse amplitudes, velocities, waiting times and durations at all radial positions.

In the reference case where all pulses have the same constant velocity, the mean value of the process decreases exponentially with radial position with an e-folding length given by the product of the pulse velocity and the linear damping time, $\Phiave(x)=\gammastar\aoave\exp{(-x/\voave\taup)}$. Here the intermittency parameter $\gammastar=\taustar/\woave$ is the ratio of the pulse duration $\taustar=\taup\ell/(\voave\taup+\ell)$ and the average waiting time $\woave$, so longer durations, higher rate and larger amplitudes of pulses increases the mean value of the process. The relative fluctuation level is given by $\Phirms/\Phiave=1/\gammastar^{1/2}$, the skewness moment is $S_\Phi=2/\gammastar^{1/2}$ and the flatness moment is $F_\Phi=6/\gammastar$. Thus, the fluctuations become stronger and more intermittent when the degree of pulse overlap decreases, that is, for shorter, faster and less frequent pulses. These normalized higher order moments are radially constant in the reference case. The analysis presented here, specifically the radial profiles presented in \Figsref{fig.prof_timeindep_Paexp} and \ref{fig:prof_exp}, describe how a distribution of pulse velocities and correlations with the amplitudes modify the moments with respect to the reference case.

The stochastic modeling has here been extended to include a distribution of pulse velocities and correlations with amplitudes, in particular considering a power law scaling of the velocity with amplitude, in accordance with blob velocity scaling theories. General expressions have been obtained for the cumulants and lowest order statistical moments, describing their radial variation due to transport along the magnetic field. Closed form expressions for the distribution and radial variation of pulse amplitudes, velocities, durations and waiting times have also been derived, elucidating all statistical properties of the process. In general, the correlations between pulse amplitudes and velocities make the process dominated by fast and large-amplitude pulses, which results in higher fluctuation amplitudes and stronger intermittency of the process.

In the case where the velocities depend on the instantaneous amplitude, $V\sim A^\alpha$, pulses will stagnate due to the linear damping. This makes the average waiting time between pulses increase radially outwards, thereby increasing the intermittency of the process. In the limit $\alpha\rightarrow0$, the pulse velocities are independent of the amplitudes and all pulses move with the same time-independent velocity. The radial variation of the average pulse parameters for pulse velocities given by a power law dependence on instantaneous amplitudes and an exponential amplitude distribution at the reference position is summarized in \Tabref{tab:summary}. A particularly interesting and relevant case is a linear dependence, $\alpha=1$. In this case, the radial variation of the mean value of the process is exponential with the same e-folding length $\voave\taup$ as for the reference case. Due to stagnation of slow and small-amplitude pulses, the average amplitude and velocity are the same for all radial positions. Moreover, both amplitudes and velocities have an exponential probability distribution. However, the average pulse waiting time increases exponentially with radial position due to pulse stagnation. As a result, the relative fluctuation level and the skewness and flatness moments increase radially outwards and become much higher than for the reference case, as shown in \Figref{fig:prof_exp}.

\begin{table}[tb]
    \centering
    \begin{tabular}{|c|c|c|c|c|}
    \hline
         & General $\alpha$ & $\alpha=0$ & $\alpha=1/2$ & $\alpha=1$
         \\
        \hline
         \Phiave & \Eqref{eq.Phiave} & $\gammastar\aoave\exp{(-x/\voave\taup)}$ & \Eqref{eq:Phiave-a=1/2} & $\gammastar\aoave\exp{(-x/\voave\taup)}$
         \\
         $\Aave$ & $-$ & $\aoave\exp{(-x/\voave\taup)}$ & \Eqref{eq:Aavea=1/2} & $\aoave$
         \\
         $\Vave$ & \Eqref{eq:Vave} & $\voave$ & \Eqref{eq:Vavea=1/2} & $\voave$
         \\
         $\wave$ & \Eqref{eq.tauwxi} & $\woave$ & $\woave\exp{((\pi^{1/2} x/2\voave\taup)^{1/2})}$& $\woave\exp{(x/\voave\taup)}$
         \\
         \hline
    \end{tabular}
    \caption{Summary of radial variation of pulse parameters for pulse velocities given by a power law dependence on instantaneous amplitudes and an exponential amplitude distribution at the reference position for various scaling exponents.}
    \label{tab:summary}
\end{table}

The radial variation of the lowest order statistical moments of the process are remarkably similar for the time-independent and time-dependent cases, presented respectively in \Figsref{fig.prof_timeindep_Paexp} and \ref{fig:prof_exp}, notably with the same plot ranges. The radial variation of the mean value of the process is close to exponential with the same e-folding length as for the reference case where all pulses have the same velocity. Furthermore, the relative fluctuation level as well as the skewness and flatness moments all increase radially outwards and become much higher than in the reference case. However, these two cases differ significantly in other respects. In particular, the rate of pulses is radially constant in the time-independent case while the average waiting time increases radially outwards for time-dependent pulses. Moreover, for $\alpha=1$, the average pulse amplitude is radially constant for the time-dependent case, while it decreases exponentially with radius for time-independent pulse velocities. The underlying dynamics is therefore markedly different for time-independent and time-dependent pulse velocities.

The results presented here have profound influences for the analysis of realizations of the process, such as experimental measurement data and first principles-based turbulence simulations of the boundary region of magnetically confined plasmas. A broad distribution of pulse velocities implies a distribution of durations, which strongly influences the auto-correlation function and frequency power spectral density of the process.\cite{garcia_auto-correlation_2017,korzeniowska_apparent_2023,korzeniowska_longrange_2024} Thus, in order to reliably estimate the model parameters from realizations, dedicated methodology must be developed.\cite{losada_three-point_2024,losada_time_2024} Conditional averaging is commonly used to reveal the pulse function and duration as well as the distribution of amplitudes and waiting times for large-amplitude fluctuations.\cite{garcia_burst_2013,garcia_intermittent_2013,kube_fluctuation_2016,theodorsen_scrape-off_2016,garcia_sol_2017,johnsen_conditional_1987,pecseli_statistical_1989,oynes_fluctuations_1995,nielsen_turbulent_1996,grulke_experimental_1999,block_prospects_2006,kube_fluctuation_2016} Recently, a deconvolution algorithm has been developed that takes the pulse function and duration as input and identifies the pulse arrivals and amplitudes from realizations of the process.\cite{theodorsen_universality_2018,ahmed_reconstruction_2023} However, the usefulness of these methods requires a systematic investigation of their performance for a broad distribution of pulse durations and various degrees of overlap. This will be presented in separate studies along with applications to experimental measurement data and turbulence simulations.

\section{Conclusions}\label{sec.conclusions}

The presence of blob-like filament structures in the scrape-off layer of magnetically confined fusion plasmas has significant implications for plasma-wall interactions and exhaust physics in fusion reactors. These filaments, which are coherent structures of plasma that move radially outward from the core plasma, strongly influence both particle and heat transport at the plasma edge. Understanding their behavior is crucial for designing and operating fusion reactors, where the boundary plasma and plasma-facing components are critical to reactor performance and longevity. The localized particle and heat fluxes on plasma-facing component can lead to issues like thermal fatigue, erosion, impurity influx, and challenges in predicting power deposition in the divertor region. As a result, future fusion reactor designs must incorporate strategies to manage these intermittent heat loads, improve material resilience to thermal fatigue, and develop more accurate models for predicting edge plasma behavior, including the dynamics of blob transport. It is therefore clear that a statistical description of the fluctuations is required, identifying the distribution of and correlation between blob amplitudes, sizes, velocities and rate of occurrence.

Single-point measurements in the boundary region of magnetically confined plasmas have demonstrated that the large-amplitude fluctuations due blob-like structures can be described as exponential pulses that appear in accordance with a Poisson process and an exponential amplitude distribution. Based on this, a statistical framework has been developed, describing the fluctuations as a super-position of uncorrelated pulses. As suggested by blob velocity scaling theories, an important feature is a strong correlation between pulse amplitudes and velocities. Moreover, the pulse velocities will depend on the instantaneous amplitude, which decreases over time due to transport along magnetic field lines. In this work, the implications of such correlations and time-dependence has been investigated in detail. A particularly interesting case is that of a linear dependence of pulse velocities on the instantaneous velocities, resulting in a radially constant mean value of the pulse amplitude and velocity while the average waiting time increases exponentially with radius. The main results are that an increasing average pulse velocity results in a flattened radial profile of the mean value of the process as well as a larger relative fluctuation level and stronger intermittency. The stochastic modeling provides numerous predictions for the distribution of pulse parameters and their radial variation that will be confronted with experimental measurements and first principles-based turbulence simulations. Such analysis will be crucial for identifying relevant validation metrics for simulation codes focused on the boundary region.

\appendix

\section{Stationarity of the process}\label{app.stat}

A pulse with a time-dependent velocity given by \Eqref{eq.velocity} will arrive at the radial position $\xi$ at time $s_\xi$, where
\begin{equation}
    s_\xi=s_0-\frac{\taup}{\alpha}\ln \left(1-\frac{\xi}{\Xmax}\right) .
\end{equation}
The pulse will stop moving as the stagnation point $\Xmax$ is approached, where $\Xmax$ is given by \Eqref{eq.Xmax}. The arrivals $s_0$ at the reference position $\xi=0$ are assumed to be uniformly distributed on the interval $[-T/2, T/2]$, that is, $P_{s_0}=1/T$. In the case of a random distribution of pulse amplitudes $a_0$ at the reference position, the pulse arrival times at the radial position $\xi$ are given by a sum of two random variables, and the distribution of these arrivals is therefore given by the convolution
\begin{equation}
    P_{s_\xi}(s)=\int\limits_{-\infty}^{\infty} \rmd r\, P_{r_\xi}(r)P_{s_0}(s-r)=\frac{1}{T} \int\limits_{\max(0, -T/2+s)}^{T/2+s} \rmd r\, P_{r_\xi}(r) .
\end{equation}
Here $P_{r_\xi}$ is the distribution of the transit times $r_\xi$ at the position $\xi$, given by $r_\xi=-(\taup/\alpha)\ln(1-\xi/\Xmax)$. For a general pulse amplitude distribution $P_{a_0}$ at the reference position, the transit time distribution is for $r_\xi>0$ given by 
\begin{equation}
        P_{r_\xi}(r) = 
        \frac{\aoave}{\taup} \frac{\left(\frac{\xi\alpha}{c_v\voave\taup}\right)^{\frac{1}{\alpha}}\exp\left(-\frac{\alpha r}{\taup}\right)}{\left(1-\exp\left(-\frac{\alpha r}{\taup}\right)\right)^{1+\frac{1}{\alpha}}}   P_{a_0}\left(\aoave\left(\frac{\xi\alpha}{c_v\voave\taup}\right)^{\frac{1}{\alpha}} \left[\left(1-\exp\left(-\frac{\alpha r}{\taup}\right)\right)^{-\frac{1}{\alpha}}-1\right]\right) .
\end{equation}
It follows that the amplitude distribution $P_{a_0}$ influences the arrival time distribution and leads to end effects in realizations of the process.\cite{losada_stochastic_2023}

\begin{figure}[tb]
\centering
\includegraphics[width=8cm]{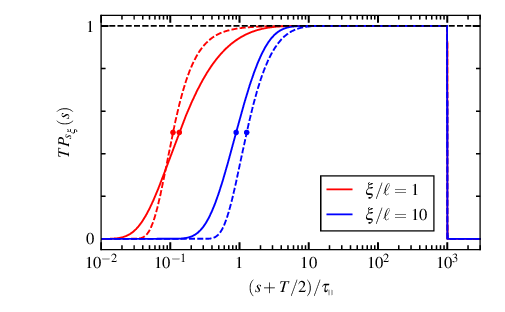}
\caption{Probability distribution of pulse arrival times $s_\xi$ for $\alpha=1/2$ (broken lines) and $\alpha=1$ (full lines) at various radial positions. The filled circles indicates $1/2T$ probability for arrival times equal to $-T/2 + \rmedian$. The normalized linear damping time is $\voave\taup/\ell=10$ and $T/\taup = 10^3$.}
\label{fig.arrivals_dist}
\end{figure}

In the reference case of an exponential distribution of pulse amplitudes at the reference position, the distribution of pulse arrival time can be calculated in closed form,
\begin{align}\label{eq.Psxi}
    TP_{s_\xi}(s) = 
    \begin{cases}
        \exp \left(-\left(\frac{\alpha\xi}{c_v\voave\taup}\right)^{\frac{1}{\alpha}}\left[ \left(1-\exp\left(-\frac{\alpha(s+T/2)}{\taup}\right)\right)^{-\frac{1}{\alpha}} - 1 \right]\right), & -T/2<s<T/2 , \\
        \exp \left(-\left(\frac{\alpha\xi}{c_v\voave\taup}\right)^{\frac{1}{\alpha}}\left[ \left(1-\exp\left(-\frac{\alpha(s+T/2)}{\taup}\right)\right)^{-\frac{1}{\alpha}} - 1 \right]\right)  &     \\
        - \exp \left(-\left(\frac{\alpha\xi}{c_v\voave\taup}\right)^{\frac{1}{\alpha}}\left[ \left(1-\exp\left(-\frac{\alpha(s-T/2)}{\taup}\right)\right)^{-\frac{1}{\alpha}} - 1 \right]\right), & T/2<s.
    \end{cases}
\end{align}
This distribution is presented in \Figref{fig.arrivals_dist} for $\alpha=1/2$ and $\alpha=1$ at two different radial positions. In the limit $\alpha\to0$, the distribution of arrival times becomes uniform, $P_{s_\xi}(s)=1/T$, over the range from $-T/2+\xi/\voave$ to $T/2 + \xi/\voave$, as discussed in detail in Ref.~\onlinecite{losada_stochastic_2023}. In general, for $\alpha>0$ there is no interval in which the arrival times are exactly uniformly distributed. Equation~\eqref{eq.Psxi} shows that there are several temporal scales involved. Firstly, there is a transient, highly non-stationary period for the pulses to arrive at position $\xi$. For the present case, this is most easily quantified by the median $\rmedian$ of the transit times, defined by
\begin{equation}
    \int_{0}^{\rmedian} \text{d}r_\xi\,P_{r_\xi}(r_\xi) = \frac{1}{2} .
\end{equation}
For the transit time distribution given by \Eqref{eq.Psxi} this gives
\begin{equation}
    \rmedian =-\frac{\taup}{\alpha} \ln \left(1-\left[1+\ln{2}\left(\frac{\alpha\xi}{c_v\voave\taup}\right)^{-\frac{1}{\alpha}}\right]^{-\alpha}\right) .
\end{equation}
This median arrival time is indicated by the filled circles in \Figref{fig.arrivals_dist}. Secondly, there is an exponential saturation given only by the linear damping time for $s+T/2\gg\taup$. In this limit, the arrival time distribution becomes arbitrarily close to a uniform distribution, as shown in \Figref{fig.arrivals_dist}.

\begin{figure}[tb]
\centering
\includegraphics[width=8cm]{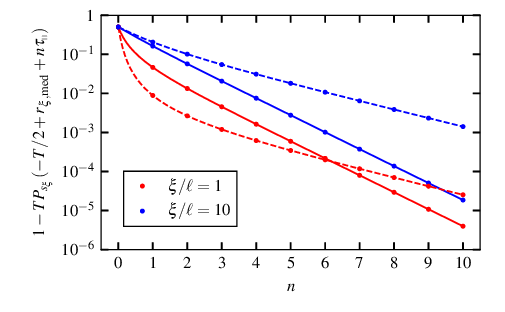}
\caption{Probability distribution of pulse arrival times for arrival times equal to $r_{\xi, \text{med}}$ plus a multiple $n$ of the linear damping time, showing effective distance from the stationary phase depending on the multiple $n$ for various radial positions. The broken lines indicate the case $\alpha=1/2$ and full lines the case $\alpha=1$. The normalized linear damping time is $\voave\taup/\ell=10$.}
\label{fig.convergence_to_stationarity}
\end{figure}

To quantify the proximity to a uniform distribution, consider the value of the arrival time distribution at the median time plus an integer multiple $n$ of the linear damping time.
The difference between the probability distribution for arrival times equal to $\rmedian+n\taup$ from the uniform distribution $1/T$ is presented in \Figref{fig.convergence_to_stationarity} for $\alpha=1/2$ and $\alpha=1$ at two different radial positions. The difference for a large multiple of the linear damping time decreases exponentially with $n$ with e-folding factor $1/\alpha$. This describes how fast the arrival time probability density approaches the uniform distribution after the median arrival time has been reached.

Assuming that $T>r_{\xi, \text{med}} + n\taup$ and $n$ is large, the pulse arrivals at the radial position $\xi$ becomes approximately uniformly distributed on the interval $[-T/2+r_{\xi, \text{med}} + n\taup, T/2]$, and hence will closely arrive according to a Poisson process with a radially varying rate $1/\tauwxi$, given by \Eqref{eq.tauwxi}. For a sufficiently long realization of the process, $T\gg r_{\xi, \text{med}} + n\taup$, end effects can be neglected and the process can be considered a stationary Poisson process in the radial domain of interest.

\section*{Acknowledgements}

This work was supported by the UiT Aurora Centre Program, UiT The Arctic University of Norway (2020). A.T.\ was supported by the Troms{\o} Research Foundation under Grant No.\ 19\_SG\_AT.

\bibliographystyle{custom}
\bibliography{SOL}

\end{document}